\documentclass[jpb,preprint]{revtex4}%
\usepackage{amsfonts}
\usepackage{amsmath}
\usepackage{amssymb}
\usepackage{graphicx}%
\begin{document}

%\pdfoutput=1

\title{Constraints on the detection of topological charge of optical vortices using
self-reference interferometry}
\author{Siyao Wu}
\affiliation{School of Mathematics and Physics, China University of Geosciences, Wuhan
430074, China}
\author{Ling Chen}
\email{lingchen@cug.edu.cn}
\affiliation{School of Mathematics and Physics, China University of Geosciences, Wuhan
430074, China}
\author{Ruiping Jing}
\affiliation{School of Mathematics and Physics, China University of Geosciences, Wuhan
430074, China}
\author{Baocheng Zhang}
\email{zhangbaocheng@cug.edu.cn}
\affiliation{School of Mathematics and Physics, China University of Geosciences, Wuhan
430074, China}

\begin{abstract}
Self-reference interferometry of optical vortices using a Michelson
interferometer is investigated in this paper. It is found that the detection
of topological charge (TC) for the optical vortices is constrained by some
physical conditions. We present these conditions through the theoretical
analyses, numerical simulation and experimental results. For different
parameters, the maximal detectable TCs are different, which is helpful for the
measurement of TC in the practical application. Within the range allowed by
the constrained conditions, we also study the detection of TC using the
interference pattern of two-way optical vortex by changing the inclined angle
of one mirror of the Michelson interferometer.

\end{abstract}

\maketitle

\section{Introduction}

Since optical vortices were found \cite{CGR1989,SGY2019}, their structure and
property such as cylindrical symmetry, dark core at the center and so on in
the propagation \cite{AG2001,DP2002} has been studied over the past few
decades. The wave vector of the optical vortex rotates around the vortex
center, which leads to an orbital angular momentum (OAM) appearing in the
phase as $exp(im\theta)$, where $m$ is topological charge (TC) \cite{LA1992}
and determines the OAM value ($m\hbar$) of the photon. The most familiar types
of optical vortices are Laguerre-Gaussian (LG) beams, Bessel and
Bessel-Guassian beams, ring Gaussian beams and hypergeometric Gassian beams,
etc. They can be generated artificially by different methods, such as the
geometrical optics model transformation method \cite{MWB1993}, the spiral
phase plate method \cite{MWB1994}, the computer generated holographic method
(CGH) \cite{NRH1992}, the spatial liquid-crystal modulation method (SLM)
\cite{JEC2002}, and the metamaterial vortex producing method
\cite{DV2015,JZ2016} etc.

Optical vortices have been applied to many different situations, such as
optical trapping \cite{KTG1996, LL2018}, optical communication
\cite{AM2001,BN2018}, microscopy imaging \cite{BS2008,AS2018}, laser
micromachines \cite{WC2018,CH2010}, rotating velocity detection
\cite{MPJ2013,BL2020}, optical analogue for black hole \cite{FM2008,DV2018},
and so on. In all these applications, TC or OAM of the optical vortex plays a
key role, so it is very important to find out the value of TC.

In terms of measuring the TC, various techniques have been proposed, such as
the diffraction grating method \cite{DF2015,SZ2017}, Doppler analysis
\cite{MVV2003}, the metamaterial surface method \cite{NY2014,JJ2016},
interferometry \cite{HIS2006,GGL2018,BK2017} and so on. Different methods have
their advantages and disadvantages. For example, in the interferometric
method, the TC can be specified by the interference of a reference beam (such
as a plane or spherical beam) and a detected vortex beam, as usually carried
out in Mach-Zehnder interferometers \cite{HLS2012,KN2019,GMS2019,GMZ2021}.
Since the same frequencies in the interferometry are required for the two
beams, the self-reference interferometry is advantageous to realize this
point. Based on the interference resulting from splitting incident wave by
Wollaston prism or birefringent plate, the optical vortex sign was determined
by self-interference methods \cite{KBM2010}. Recently, B. Lan et al\textit{.}
carried out the dislocation self-reference interferometry based on Michelson
interferometer to measure the TC of the vortex beam \cite{BL2019}. In their
method, the TC can be measured accurately and the additional reference beam is
not required. However, they didn't realize the measurement limits of this
technique, which provides the motivation for our work in this paper. We will
show that the measurable maximal value of TC using self-reference
interferometry is affected by some physical parameters such as the radius of
the beam and so on.

In this paper, we investigate the self-reference interferometry \cite{BL2019}
through the theoretical and experimental avenues using a Michelson
interferometer and discuss the measurement limits of this method. It is noted
that the interference stripes will become fat when centers of the two vortex
beams approach to each other, which limits the measurable values of the TC and
has not been observed before. For presenting this, the paper is organized as
followed. In the second section, we study the constraint on the detection for
the TC of the optical vortex by theoretical analyses, numerical simulation and
experimental demonstration. A fundamental constrained relation is given
through purely theoretical analysis. At the same time, the constraint from the
experimental conditions is analyzed. Within the allowed range by the
constrained conditions, we discuss three different cases for the
self-reference interference, and compare the simulated and experimental
results in the third section. Meanwhile, we also discuss the capacity using
the self-reference interferometry to detect whether two beams are parallel
(whether the mirrors in the Michelson interferometer are vertical). Finally,
we give the conclusion in the fourth section.

\section{Constraint on self-reference interferometry}

\begin{figure}[ptb]
\centering
\includegraphics[width=1\columnwidth]{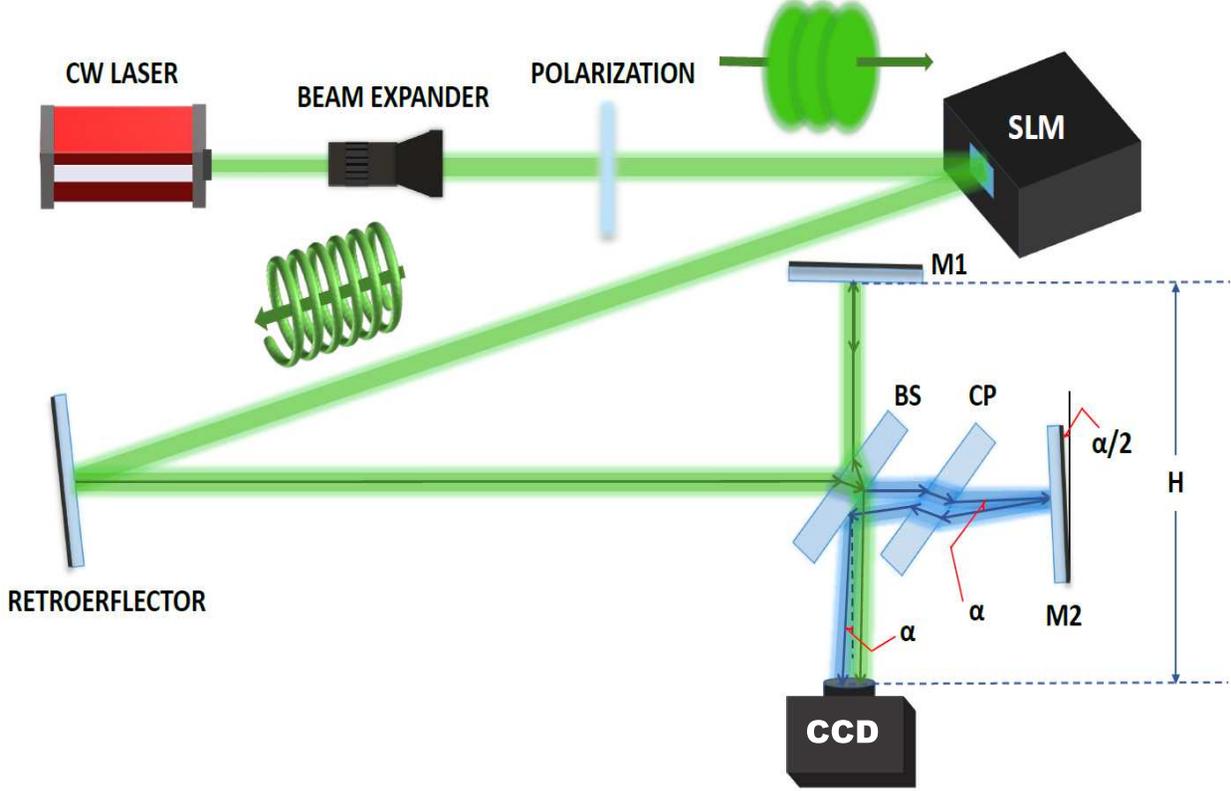}\caption{Experimental setup. A
CW 532 nm laser beam is modulated to be a Laguerre-Gauss beam by a SLM before
being launched into the Michelson interferometer. Green and blue optical
pathes are reflected by a fixed mirror $M1$ and a moving mirror $M2$. The
interference pattern is imaged onto a cold CCD camera after the green light
goes through the beam splitter (BS) and the blue light goes through the
compensation plate (CP) and the BS. The inclined angle of the optical path is
controlled by the moving mirror $M2$ by adjusting two screws in the device.}
\label{Fig1}
\end{figure}

We start with two vortex beams, one of which goes along the optic axis and
another one goes with an inclined angle $\alpha$ (for its practical meaning,
see the experimental setup in Fig. 1 which will be discussed later) from the
optic axis. They can be described with the complex optical fields $E_{OAM1}$
and $E_{OAM2}$ as%
\begin{equation}
E_{OAM1}=A_{1}exp(im_{1}\theta),
\end{equation}%
\begin{equation}
E_{OAM2}=A_{2}exp(im_{2}\theta^{^{\prime}}+ikrsin\alpha),\label{2}%
\end{equation}
where $A_{1}$, $A_{2}$ and $m_{1}$, $m_{2}$ denote the amplitude and TCs of
the two vortex beams, respectively. $\theta$ and $\theta^{^{\prime}}$ are the
azimuthal angles of the vortex beams, $k$ is the wave vector, and $r$ is the
radial distance on the transverse plane perpendicular to the propagation axis.
The additional phase $ikrsin\alpha$ originates from the inclination of one
beams from another one during the propagation and the value is obtained by
projecting the inclined wave vector on the transverse plane. This term
$ikrsin\alpha$ is equivalent to that term called as $\varphi_{carry}$ in Ref.
\cite{BL2019}, but can present unambiguously the inclined angle here. When the
two vortex beams are received by the charge-coupled device (CCD), the
interference pattern would be presented according to the following way given
by the distribution of the total light intensity,%
\begin{equation}
I=|E_{OAM1}+E_{OAM2}|^{2}=A_{1}^{2}+A_{2}^{2}+2A_{1}A_{2}cos[(m_{2}%
\theta^{^{\prime}}-m_{1}\theta)+krsin\alpha],\label{3}%
\end{equation}
where $(m_{2}\theta^{^{\prime}}-m_{1}\theta)$ is the phase difference caused
by the coherent two vortex beams, which can be read from the interference
pattern by the number and the direction of the branches. As the vortex beam is
considered as self-coherent in our paper, the TCs and amplitudes for two
vortex beams should be the same, i.e. $m_{1}=m_{2}=m$ and $A_{1}=A_{2}=A$.
Thus, the intensity distribution becomes%
\begin{equation}
I=|E_{OAM1}+E_{OAM2}|^{2}=2A^{2}+2A^{2}cos[m(\theta^{^{\prime}}-\theta
)+krsin\alpha],\label{4}%
\end{equation}

\begin{figure}[ptb]
\centering
\includegraphics[width=1\columnwidth]{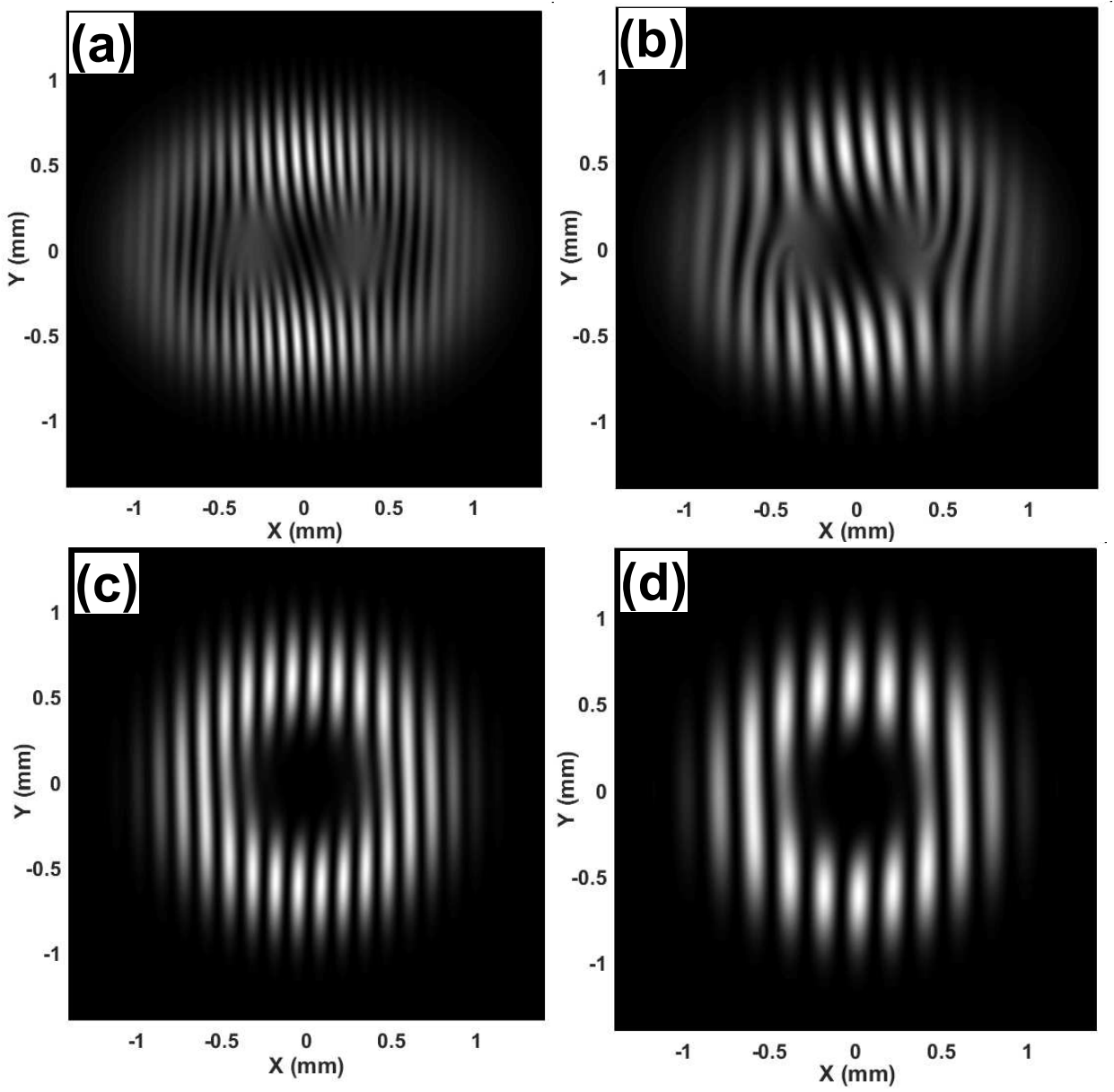}\caption{Simulated results for
the interference patterns with the different distance between the centers of
two vortex beam: (a) $d=0.60 mm$; (b) $d=0.36 mm$; (c) $d=0.24 mm$; (d)
$d=0.12 mm$. The value of TC for the vortices is taken as $m=3$ for the
simulation. }
\label{Fig2}
\end{figure}

At first, we simulate the interference using two LG vortex beams. It is noted
that the width of the interference stripes will increase when the distance $d$
(that is controlled by the inclined angle $\alpha$ in our paper) between the
centers of the two vortex beams decreases, as presented in Fig. 2. The
branching number at each fork point denotes the number $m$ of TC, and thus,
the TC of the vortex beam can be detected by observing the number of the
branches. Thus, the TC cannot be read from the Fig. 2 (d) in which the stripes
is so fat that the number of branches is covered at the fork points. This is a
new phenomena which is not observed before. In particular, it gives a limit
for the measurement of the TC, and the related interpretation will be given below.

The existence of the measurement limit stems from two reason. One is
fundamental which is limited by the distance $d$ between the centers of two
optical vortices or the inclined angle $\alpha$. Another one is experimental
which is limited by the pixel of the detector. In order to explain this
clearly, we have to state the experimental setup to make the parameters be
understood unambiguously.

In our experimental setup, a Michelson interferometer is used for the
realization of the self-reference interferometry, and the adjustment of the
mirrors can lead to the parallel and non-parallel optical paths. The practical
instruments and optical paths are shown in Fig. 1. The continuous-wave (CW) of
a Gauss laser beam is emitted from the semiconductor laser (CNI, China) with
the mode $TEM_{00}$ and the wavelength $532nm$. The diameter of the beam is
expanded to $2mm$ by a beam expander. Then, it is modulated to be a typical
vortex beam, a Laguerre-Gauss beam with the radial mode number $p=0$ and TC
$m=3$ by a spatial light modulator (SLM, UPLabs, China). The conversion
efficiency (or the zero-order diffraction efficiency) of the SLM for
modulating the $TEM_{00}$ beam to $LG_{03}$ beam is about $80\%$. The
generated vortex beam is pumped to the Michelson interferometer and is split
into two coherent vortex beams which self-interfere on the imaging plane of a
CCD. The interference pattern is produced in the cold CCD (Tucsen, China) with
an array of 1912*1452 pixels (the size of each one is about 4.5 $\mu m$). In
Michelson interferometer, one of the optical path is reflected by a fixed
mirror $M1$, and another optical path is reflected by a moving mirror $M2$. We
can easily adjust two screws of $M2$ to change the inclined angle of the
second optical path while the first path is fixed as the optical axis. The
inclined angle $\alpha$ and the distance $H$ between the imaging plane of the
CCD camera and the mirror $M1$ (note that two armlengths are equal in our
experiment) are marked in Fig. 1.

We firstly interpret the limit caused by the distance between the centers of
two vortex beams or by the inclined angle. When the self-interference happens,
the branching could appear in the interference pattern. In order to count the
TC by the number of the branches at the fork points, there are $(m+2)$
discernable fringes at least in the region between the centers of two vortex
beams, which is guaranteed by the relation,%
\begin{equation}
d\geqslant\left(  m+2\right)  \Delta d.\label{dtv}%
\end{equation}
where $\Delta d=\lambda/\sin(\alpha)$ is the distance between two nearby
bright fringes and the width for the bright and dark fringes is equal. Given
the distance $H$, the distance $d$ can be gotten using the inclined angle
$\alpha$ as $d=H\tan(\alpha)$. Substitute this expression into the relation
(\ref{dtv}) and obtain the constraint for the detectable TC as,%
\begin{equation}
m\leqslant\frac{H}{\lambda}\tan(\alpha)\sin(\alpha)-2.\label{ctc}%
\end{equation}
This shows that the larger the value of the inclined angle is, the larger the
number of TC allowed to be measured is\textit{. }At the same time, the
distance $d$ must be constrained to ensure that the singular points for the
phase cannot exceed the edge of the intensity-overlapped region of two vortex
beams, which requires
\begin{equation}
d=H\tan(\alpha)\leq r_{0},\label{orv}%
\end{equation}
where $r_{0}$ is the radius of each beam. Compare the two relations
(\ref{ctc}) and (\ref{orv}), and we obtain a new relation to constrain the
value of detectable TC,%
\begin{equation}
m\leqslant\frac{r_{0}\sin(\arctan(r_{0}/H))}{\lambda}-2.\label{ocr}%
\end{equation}
This is our central result. It shows that the maximal detectable TC ($m_{\max
}=\frac{r_{0}\sin(\arctan(r_{0}/H))}{\lambda}-2$) is modulated by the radius
of the beam and the distance $H$. Note that $m_{\max}$ has the same changing
trend with the radius of the beam as seen from the left plot of Fig. 3, but
has the inverse changing trend with the distance $H$ as seen from the right
plot of Fig. 3. In our experiment, $\lambda=532nm$, $H=30cm$, and $r_{0}=1mm$,
the constraint (\ref{ocr}) gives $m\leqslant4$, consistent with our
experimental presentation. Larger values for TC can also be detected if the
radius $r_{0}$ is increased or the distance $H$ is decreased. If the radius
$r_{0}$ increases to $1cm$ and the distance $H$ decreases to $10cm$, the
detectable TC can approach to $1880$, which is possible but depends on the
experimental conditions such as the pixel of the CCD camera and so on.

\begin{figure}[ptb]
\centering
\includegraphics[width=0.5\columnwidth]{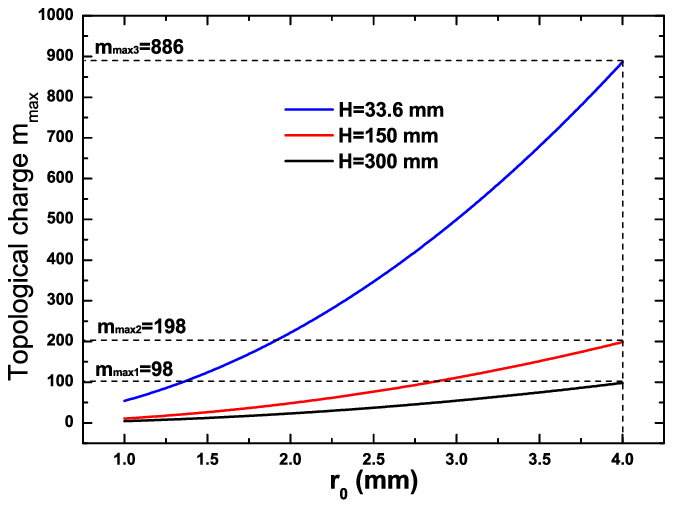}\includegraphics[width=0.5\columnwidth]{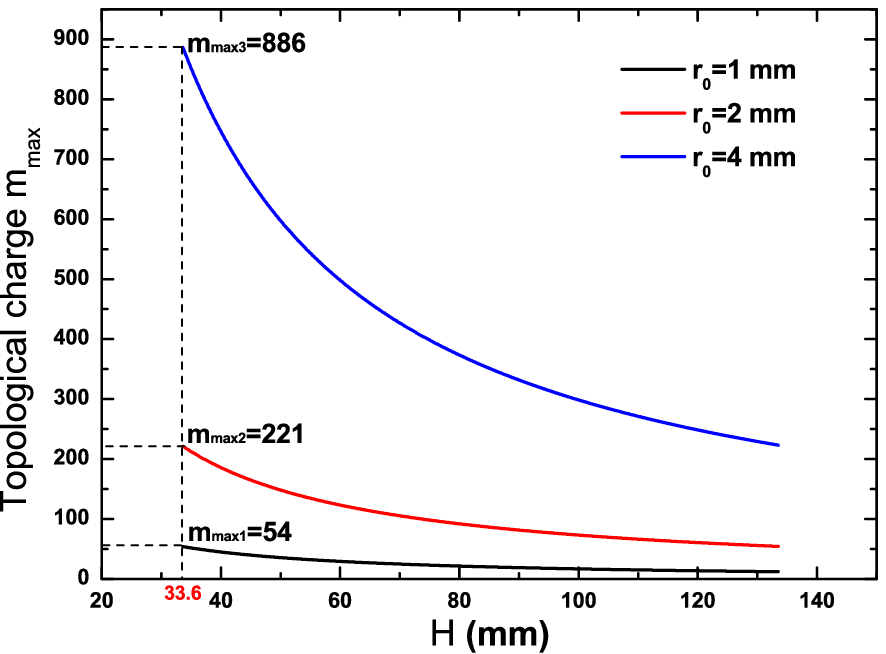}\caption{Maximal
detectable TC as a function of parameters $r_{0}$ and $H$. The left plot
describes the change of maximal TC with $r_{0}$ for three different fixed $H$,
and for every $H$, the value of maximal TC is marked for the experimentally
allowed largest value of $r_{0}$. The right plot describes the change of
maximal TC with $H$ for three different fixed $r_{0}$, and for every $r_{0}$,
the value of maximal TC is marked for the experimentally smallest value of
$H$.}
\label{Fig3}
\end{figure}

Then, we discuss the limit from the experimental conditions and focused on the
influence from the pixel of the CCD camera. As discussed above, the distance
$\Delta d$ between two nearby bright fringes is $\Delta d=\lambda/\sin
(\alpha)$. Thus, the spatial frequency of the fringes for one pixel
$d_{pixel}$ can be expressed as $f=d_{pixel}/\Delta d=d_{pixel}sin(\alpha
)/\lambda$, which means that when the observation is made at the imaging plane
of CCD camera, the number of fringes observed at each pixel is $f$. In order
to distinguish the fringes clearly, the number of the fringes appeared on CCD
should be equal to that obtained through the interference, which means that a
single fringe should be converted on a single pixel of CCD. Mathematically, it
requires $f\leqslant1$ or%
\begin{equation}
\alpha\leqslant\arcsin\left(  \frac{\lambda}{d_{pixel}}\right)  . \label{iac}%
\end{equation}
This shows the geometric parameter $\alpha$ is constrained by the experimental
condition and cannot be increased or decreased at random. In our experiment,
$\lambda=532nm$, $d_{pixel}=4.5\mu m$ which gives that the inclined angle
cannot be more than $\alpha_{\max}\simeq0.118$.

According to our CCD, the radius of the beam can only increase to $r_{0}=4mm$
at most which demands%
\begin{equation}
H\geqslant\frac{r_{0}}{\tan(\alpha_{\max})}\simeq33.6mm, \label{dcr}%
\end{equation}
where the relations (\ref{orv}) and (\ref{iac}) are used. Together with the
relation (\ref{ocr}), it shows that the distance $H$ cannot be decreased at
random and the radius $r_{0}$ cannot be increased at random for the purpose of
detecting large TC, and they must satisfy such relation as (\ref{dcr}) that
present the experimental constraint. The maximal TC that can be detected is
$886$ by substituting $H=33.6mm$ and $r_{0}=4mm$ into the relation
(\ref{ocr}), but in the practical experimental implementation, the distance
$H$ cannot take so small value.

We also present experimentally the limit of measuring TC by self-reference
interferometry using the setup in Fig. 1. The experiment begins from so far
distance between two vortex beams that the two beams are not overlapped, as
given in Fig. 4(a), to the distance where no branching can be observed, as
given in Fig. 4(h). It is seen that the stripes become fatter as the two
vortex beams approach to each other, which means that the experimental
measurement of TC will be limited when the stripes becomes so fat that the two
fork points cannot be distinguished, as shown in Fig. 4 (g) and (h). In
particular, the interference at the least distance $d$ (about $0.6mm$) where
the branching just cannot be observed is presented in Fig. 4 (g).

\begin{figure}[ptb]
\centering
\includegraphics[width=1\columnwidth]{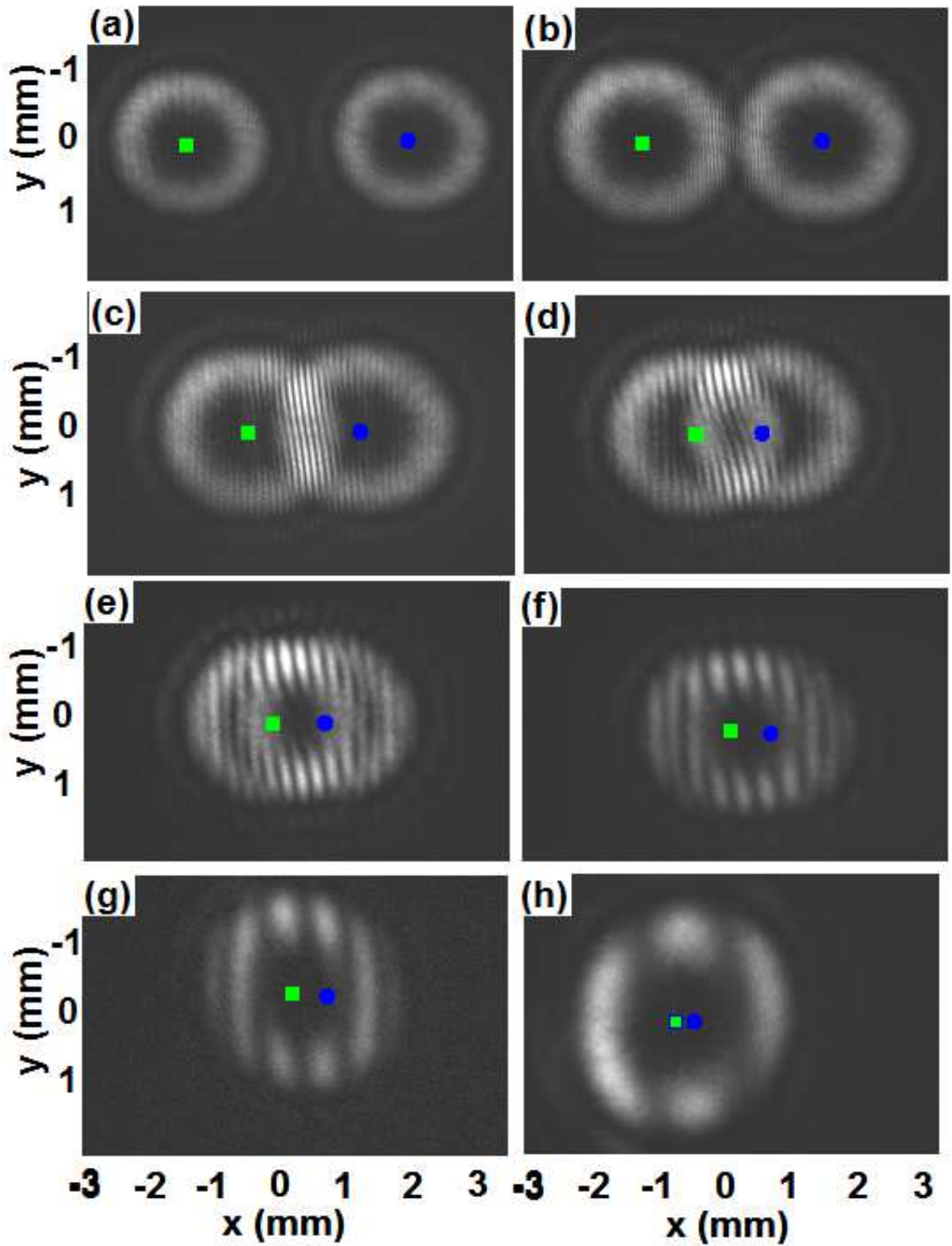}\caption{Experimental results
for the interference patterns of two vortex beams using self-reference method
when the distance between centers of two vortex beams decreases gradually and
uniformly from $3.2 mm$ to $0.3 mm$, with $m=3$ for the value of TC of the
vortices .}
\label{Fig4}
\end{figure}

\section{Measurement of topological charges}

As discussed above, the self-reference interferometry has a constraint for the
detection of TC. Only if the detection is allowed and does not exceed the
range of the constraint, this method is still nice. In this section, we will
present the detection of TC for several different cases.

In Fig. 5, we present three different cases for the propagation of the two
vortex beams. For the Fig. 5 (a), the two beams propagates along the parallel
paths, but the optical centers are overlapped partly. For the Fig. 5 (b) and
(c), the two beams propagates non-parallelly, in which the blue path is
inclined from the green one at the angle $\alpha$. In particular, the optical
centers are overlapped completely at the received place in Fig. 5 (b). All the
three cases have the overlapped parts for the two vortex beams to ensure the
interference can occur.

\begin{figure}[ptb]
\centering
\includegraphics[width=1\columnwidth]{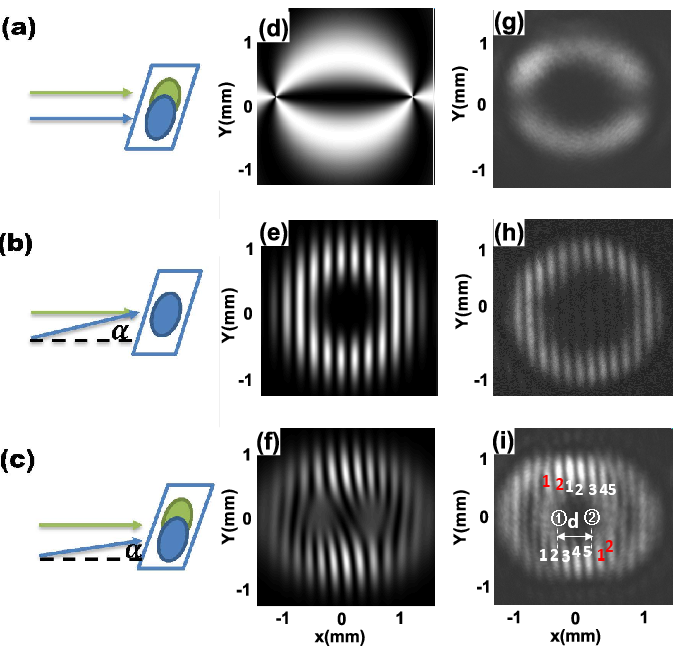}\caption{Schematic diagram of
optical paths, simulated and experimental results for different cases. (a)-(c)
are the different cases for the two optical pathes, their corresponding
interference patterns are presented in (d)-(f) using numerical method, and
their corresponding experimental results are given in (g)-(i). The blue and
green arrow lines represent two vortex beams. $\alpha$ is the inclined angle
of one of the vortex beams. The value of TC for the vortices is $m=3$ in these
figures.}
\label{Fig5}
\end{figure}

When two optical paths are parallel ($\alpha=0$) but the optical centers of
the two vortex beams are misaligned ($\theta^{^{\prime}}\neq\theta$), the
interference pattern appears according to the distribution of the light
intensity
\begin{equation}
I=|E_{OAM1}+E_{OAM2}|^{2}=2A^{2}+2A^{2}cos[m(\theta^{^{\prime}}-\theta)],
\label{5}%
\end{equation}
which derives from Eq. (\ref{4}) with the zero inclined angle. The
interference pattern looks like two arcs connecting two singularities shown in
Fig. 5 (d) numerically and Fig. 5(g) experimentally.

For the case presented in Fig. 5 (b) ($\alpha\neq0,\theta^{^{\prime}}=\theta
$), two phase singularities coincide, and the interference pattern is given by
the straight stripes without a fork point, as shown in Fig. 5 (e) numerically
and Fig. 5(h) experimentally.

Considering misaligned centers with shearing interference ($\alpha\neq
0,\theta^{^{\prime}}\neq\theta$) in Fig. 5 (c), there are two phase
singularities seen from the fork points introduced by $\theta^{^{\prime}}$ and
$\theta$, respectively. The interference pattern could be interpreted with the
pattern of two opposite-directional forks as shown in Fig. 5 (f) numerically
and Fig. 5 (i) experimentally, like a \textquotedblleft
classical\textquotedblright\ interference between a vortex beam and a plane
beam. Meanwhile, the TC of the vortex beam can be detected by observing the
branching number.

Besides the three cases, there is a trivial one that two optical paths are
completely overlapped ($\alpha=0,\theta^{^{\prime}}=\theta$) which is not
plotted, since there is no pattern appearing due to the interference cancellation.

Note that we use the inclined angle to modulate the distance between the
centers of two vortex beams. Thus, we can check if the two mirrors $M1$ and
$M2$ in the Michelson interferometer are vertical based on the self-reference
interferometry. The inclination of one of the paths brings about the
transverse wave vector component to interfere with the angular wave vector
component, which leads to the change of the interference pattern. When the
inclination exists and the two paths are not parallel, the TC can be observed
evidently. When the ray of light reflected by $M1$ and the ray of light
reflected by $M2$ are strictly vertical, two vortex beams interfere without
branches. So, the appearance of the branches in the interference pattern is an
indication that the two mirrors $M1$ and $M2$ are not vertical, which provides
a precise measurement for a slight inclination of the mirror $M2$. Moreover,
different methods using the stripes of an interference pattern to infer that
the mirror $M2$ is tilted is also very well established, the interested
readers can refer to \cite{PH2010}.

\section{Conclusions}

In this paper, we investigate the self-reference interferometry based on the
Michelson interferometer, which can be used for the measurement of TC of the
optical vortex or checking whether the propagating paths of two vortex beams
are parallel. In this method, the coherent beams are derived from the same
vortex beam modified by the SLM and the inclination is controlled by adjusting
the mirror $M2$ of the interferometer. Novelly, we find that the detection of
TC must satisfy some constrained conditions. We study the conditions in detail
and obtain that the limits of measurement (the maximal values of detectable
TC) are different for the different values taken for the physical parameters
as the radius of the beam or the distance $H$. Thus, TC of the optical vortex
cannot be detected from the self-reference interferometry beyond these
constrained conditions. These were not realized before. As a practical
application, we analyze several cases theoretically and experimentally for
demonstrating the relation among the interference pattern, the inclination and
the parallelism of the optical paths within the allowed range of the
constrained conditions. This is also significant in the practical operation
using the Michelson interferometer.

\section{Acknowledgments}

We would like to the anonymous referees for their critical and helpful
suggestions and comments. This work is supported by the NSFC under Grant No. 11654001.


\section{References}

\begin{thebibliography}{99}                                                                                               %


\bibitem {CGR1989}P. Coullet, L. Gil, and F. Rocca, Optical vortices. Opt.
Commun. 73, 403 (1989).

\bibitem {SGY2019}Y. Shen, X. Wang, Z. Xie, C. Min, X. Fu, Q. Liu, M. Gong,
and X. Yuan, Optical vortices 30 years on: OAM manipulation from topological
charge to multiple singularities. Light: Science and Application 8, 90 (2019).

\bibitem {AG2001}A. Grover, Jr. Swartlander, Peering in to darkness with a
vortex spatial filter. Opt. Lett. 26, 497 (2001).

\bibitem {DP2002}D. Palacios, D. Rozas, and G. A. Swartzlander, Observed
scatteringing into a dark optical vortex core. Phys. Rev. Lett. 88, 103902 (2002).

\bibitem {LA1992}L. Allen, M. W. Beijersbergen, R. J. C. Spreeuw, and J. P.
Woerdman, Orbital Angular-Momentum of Light and the Transformation of
Laguerre-Gaussian Laser Modes. Phys. Rev. A 45, 8185 (1992).

\bibitem {MWB1993}M. W. Beijersbergen, L. Allen, H. E. L. O. van der Veen, and
J. P. Woerdman, Astigmatic laser mode converters and transfer of orbital
angular momentum. Opt. Commun. 96, 123 (1993).

\bibitem {MWB1994}M. W. Beijersbergen, R. P. C. Coerwinkel, M. Kristensen, and
J. P. Woerdman, Helical wave front laser beams produced with a spiral phase
plate. Opt. Comm. 112, 321 (1994).

\bibitem {NRH1992}N. R. Heekenberg, R. McDuff, C. P. Smith, and A. G. White,
Generation of optical phase singularities by computer-generated holograms.
Opt. Lett. 17, 221 (1992).

\bibitem {JEC2002}J. E. Curtis, B. A. Koss, D. G. Grier, Dynamic holographic
optical tweezers. Opt. Comm. 207, (2002).

\bibitem {JZ2016}J. Zeng, L. Li, X. Yang, and J. Gao, Generating and
separating twisted light by gradient-rotation split-ring antenna metasurfaces.
Nano Letters. 16, 3101 (2016).

\bibitem {DV2015}D. Veksler, E. Maguid, N. Shitrit, D. Ozeri, V. Kleiner, and
E. Hasman, Multiple wavefront shaping by metasurface based on mixed random
antenna groups. ACS Photonics 2, 661 (2015).

\bibitem {KTG1996}K. T. Gahagan and G. A. Swartzlander, Optical vortex
trapping of particles. Opt. Lett. 21, 827 (1996).

\bibitem {LL2018}L. Li, C. Chang, X. Yuan, C. Yuan, S. Feng, S. Nie, and J.
Ding, Generation of optical vortex array along arbitrary curvilinear
arrangement. Opt. Express 26, 9798 (2018).

\bibitem {AM2001}A. Mair, A. Vaziri, G. Weihs, and A. Zeilinger, Entanglement
of the orbital angular momentum states of photons. Nature 412, 313 (2001).

\bibitem {BN2018}B. Ndagano, I. Nape, M. A. Cox, C. Rosales-Guzman, and A.
Forbes, Creation and detection of vector vortex modes for classical and
quantum communication. J. Lightwave Technol. 36, 292 (2018).

\bibitem {BS2008}B. Spektor, A. Normatov, J. Shamir, Singular beam microscopy.
Appl. Opt. 47, A78 (2008).

\bibitem {AS2018}A. Serrano-Trujillo and M. E. Anderson, Surface profilometry
using vortex beams generated with a spatial light modulator. Opt. Commun. 427,
557 (2018).

\bibitem {WC2018}W. Cheng, X. Liu, and P. Polynkin, Simultaneously spatially
and temporally focused femtosecond vortex beams for laser micromachining. J.
Opt. Soc. Am. B 35, B16 (2018).

\bibitem {CH2010}C. Hnatovsky, V. G. Shvedov, W. Krolikowski, and A. V. Rode,
Materials processing with a tightly focused femtosecond laser vortex pulse.
Opt. Lett. 35, 3417 (2010).

\bibitem {MPJ2013}M. P. J. Lavery, Detection of a spinning object using
light's orbital angular momentum. Science 341, 1175 (2013).

\bibitem {BL2020}B. Liu, H. Giddens, Y. Li, Y. He, and W.-S. Wai, Design and
experimental demonstration of Doppler cloak from spatiotemporally modulated
metamaterials based on rotational Doppler effect. Opt. Express 28, 3745 (2020).

\bibitem {FM2008}F. Marino, Acoustic black holes in a two-dimensional
\textquotedblleft photon fluid\textquotedblright. Phys. Rev. A 78, 063804 (2008).

\bibitem {DV2018}D. Vocke, C. Maitland, and A. Prain, Rotating black hole
geometries in a two-dimensional photon superfluid. Optica 5, 2334 (2018).

\bibitem {DF2015}D. Fu, D. Chen, R. Liu, Y. Wang, H. Gao, F. Li and P. Zhang,
Probing the topological charge of a vortex beam with dynamic angular double
slits. Opt. Lett. 40, 788 (2015).

\bibitem {SZ2017}S. Zheng and J. Wang, Measuring orbital angular momentum
(OAM) states of vortex beams with annular gratings. Sci. Rep. 7, 40781 (2017).

\bibitem {MVV2003}M. V. Vasnetsov, J. P. Torres, D. V. Petrov, and L. Torner,
Observation of the orbital angular momentum spectrum of a light beam. Opt.
Lett. 28, 2285 (2003).

\bibitem {NY2014}N. Yu and F. Capasso, Flat optics with designer metasurfaces.
Nat. Mater. 13, 139 (2014).

\bibitem {JJ2016}J. Jin, J. Luo, X. Zhang, H. Gao, X. Li, M. Pu, P. Gao, Z.
Zhao, and X. Luo, Generation and detection of orbital angular momentum via
metasurface. Sci. Rep. 6, 24286 (2016).

\bibitem {HIS2006}H. I. Sztul and R. R. Alfano, Double-slit interference with
Laguerre--Gaussian beams. Opt. Lett. 31, 999 (2006).

\bibitem {BK2017}B. Khajavi and E. J. Galvez, Determining topological charge
of an optical beam using a wedged optical flat. Opt. Lett. 42, 1516 (2017).

\bibitem {GGL2018}G. G. Liu, K. Wang, Y.-H. Lee, D. Wang, P. P. Li, F. Gou, Y.
Li, C. Tu, S. T. Wu, and H. T. Wang, Measurement of the topological charge and
index of vortex vector optical fields with a space-variant half-wave plate.
Opt. Lett. 43, 823 (2018).

\bibitem {HLS2012}H-C. Huang, Y.-T. Lin, and M.-F. Shih, Measuring the
fractional orbital angular momentum of a vortex light beam by cascaded
Mach--Zehnder interferometers. Opt. Comm. 285, 383 (2012).

\bibitem {KN2019}P. Kumar and N. K. Nishchal, Modified Mach--Zehnder
interferometer for determining the high-order topological charge of
Laguerre--Gaussian vortex beams. J. Opt. Soc. Am. A 36, 1447 (2019).

\bibitem {GMS2019}K. N. Gavril'eva, A. Mermoul, A. A. Sevryugin, E. V.
Shubenkova, M. Touil, I. M. Tursunov, E. A. Efremova, and V. Yu. Venediktov,
Detection of optical vortices using cyclic, rotational and reversal shearing
interferometers. Optics and Laser Technology 113, 374 (2019).

\bibitem {GMZ2021}X. Guo, Z. Meng, J. Li, J.-Z. Yang, M. Aili, and A.-N.
Zhang, The interference properties of single-photon vortex beams in
Mach--Zehnder interferometer. Appl. Phys. Lett. 119, 011103 (2021).

\bibitem {KBM2010}P. Kurzynowski, M. Borwi\'{n}ska, and J. Masajada, Optical
vortex sign determination using self-interference methods. Optica Applicata
XL(1), 165 (2010).

\bibitem {BL2019}B. Lan, C. Liu, D. Rui, M. Chen, F. Shen, and H. Xian, The
topological charge measurement of vortex beam based on dislocation
self-reference interferometry. Physica Scripta 94, 055502 (2019).

\bibitem {PH2010}P. Hariharan, Basics of Interferometry (Second Edition).
Elsevier Academic Press, Holland (2007).
\end{thebibliography}
\end{document}